\documentstyle[12pt]{article}

\begin{document}

\baselineskip=14pt plus 1pt minus 1pt

\begin{center}{\large \bf
$\Delta I=1$ staggering in octupole bands of light actinides: 
``Beat'' patterns}

\bigskip\bigskip

{Dennis Bonatsos$^{\#}$,
C.~Daskaloyannis$^+$,
S. B. Drenska$^\dagger$,
N. Karoussos$^{\#}$,
N. Minkov$^\dagger$, 
P. P. Raychev$^\dagger$,
R. P. Roussev$^\dagger$
\bigskip

{$^{\#}$ Institute of Nuclear Physics, N.C.S.R.
``Demokritos''}

{GR-15310 Aghia Paraskevi, Attiki, Greece}

{$^+$ Department of Physics, Aristotle University of
Thessaloniki}

{GR-54006 Thessaloniki, Greece }

{$^\dagger$ Institute for Nuclear Research and Nuclear Energy, Bulgarian
Academy of Sciences }

{72 Tzarigrad Road, BG-1784 Sofia, Bulgaria}}

\end{center}

\bigskip\bigskip
\centerline{\bf Abstract} \medskip
The $\Delta I=1$ staggering (odd--even staggering) in octupole bands of light 
actinides is found to exhibit a ``beat'' behaviour as a function of the 
angular momentum $I$, forcing us to revise the traditional belief that 
this staggering decreases gradually to zero and then remains at this 
zero value. Various algebraic models (spf-Interacting Boson Model, 
spdf-IBM, Vector Boson Model, Nuclear Vibron Model) are shown to predict 
in their su(3) limits
constant staggering for this case, being thus unable to describe the 
``beat'' behaviour.  
An explanation of the ``beat'' behaviour is given in terms of two 
Dunham expansions (expansions in terms of powers of $I(I+1)$~) with slightly 
different sets of coefficients for the ground state band and the negative 
parity band, the difference in the values of the coefficients being
attributed 
to Coriolis couplings to other negative parity bands. 
Similar ``beat'' patterns have already been seen in rotational bands 
of some diatomic molecules, like AgH. 

\bigskip

PACS numbers: 21.10.Re, 21.60.Fw, 21.60.Ev

Section heading: Nuclear structure 

\newpage

{\bf 1. Introduction} 

Rotational nuclear spectra have been long attributed to quadrupole
deformations
\cite{BM}, corresponding to nuclear shapes produced by the revolution 
of an ellipsis around its maximum or minimum axis and rotating around an axis 
perpendicular to their axis of symmetry. In addition, 
it has been suggested that octupole deformation occurs in certain 
regions, most notably in the light actinides \cite{Schueler} 
and in the $A\approx 150$ mass region \cite{Phillips,Sheline}, 
corresponding to pear-like nuclear shapes 
\cite{Leander,LeanderSh,Ahmad,Butler}. 
In even nuclei exhibiting octupole deformation the ground state band, which 
contains energy levels with $I^{\pi} = 0^+$, $2^+$, $4^+$, $6^+$, \dots, 
is accompanied by a negative parity band containing energy levels 
with $I^{\pi} =1^-$, $3^-$, $5^-$, $7^-$, \dots. After the first few 
values of angular momentum $I$ the two bands become interwoven, forming 
a single octupole band with levels characterized by $I^{\pi}=0^+$, 
$1^-$, $2^+$, $3^-$, $4^+$, $5^-$, \dots \cite{Schueler,Phillips,Sheline}. 
(It should be noted, however, 
that in the light actinides alternative interpretations of these 
bands in terms of alpha clustering have been proposed \cite{DaleyI,Buck}.)

It has been observed \cite{Nazar}
that in octupole bands the levels with odd $I$ and 
negative parity ($I^{\pi}=1^-$, $3^-$, $5^-$, \dots) are displaced 
relatively to the levels with even $I$ and positive parity ($I^{\pi}=0^+$, 
$2^+$, $4^+$, \dots), i.e. the odd levels do not lie at the energies 
predicted by an $E(I)=A I (I+1)$ fit to the energy levels, but 
all of them lie systematically above or all of them lie 
systematically below the predicted energies. This is an example of 
{\sl odd--even staggering} or {\sl $\Delta I=1$ staggering}, the latter 
term due to the fact that each energy level with angular momentum $I$ 
is displaced relatively to its neighbours with angular momenta $I \pm 1$. 

A similar $\Delta I=1$ staggering effect (i.e.  a relative displacement 
of the levels with odd $I$ with respect to the levels of even $I$) is known 
to occur in rotational $\gamma$ bands of even nuclei \cite{PLB200}, 
the difference being 
that in $\gamma$ bands all levels possess positive parity. 

The $\Delta I=1$ staggering effect is different from the {\sl $\Delta I=2$ 
staggering} effect recently observed \cite{Fli1,Fli2,Ced} in superdeformed 
nuclear bands \cite{Twin,Nolan,Janssens},
since the $\Delta I=2$ staggering effect refers to the systematic 
displacement of the levels with $I=2$, 6, 10, 14, \dots relatively to the 
levels with $I=0$, 4, 8, 12, \dots, i.e. in this case the level with 
angular momentum $I$ is displaced relatively to its neighbours with 
angular momenta $I\pm 2$. 

On the other hand, rotational spectra of diatomic molecules \cite{Herz}
are known to show
great similarities to nuclear rotational spectra, having in addition the 
advantage that observed rotational bands in several diatomic molecules 
are much longer than the usual rotational nuclear bands. In fact both 
$\Delta I=1$ \cite{RMD}
and $\Delta I=2$ staggering effects \cite{PRA54,PRA1999}
have been recently observed 
in rotational spectra of several diatomic molecules. $\Delta I=2$ 
staggering has been attributed \cite{PRA1999}
to the presence of one or more bandcrossings \cite{Pavli,MRM}, 
while $\Delta I=1$ staggering remains an open problem. 

It should be noted that all these effects are much larger than the relevant 
experimental errors, with the notable exception of the $\Delta I=2$ 
staggering effect in superdeformed nuclear bands \cite{Fli1,Fli2,Ced}, 
for which only one 
case (the (a) band of $^{149}$Gd \cite{Fli2}) is known to show an effect 
outside the limits of the experimental errors.  

The dependence of the amplitude of the staggering effect on the angular 
momentum $I$ presents much interest. The situation up to now has as follows:

1) Algebraic models of nuclear structure appropriate for the description 
of octupole bands, like the spf-Interacting Boson Model (spf-IBM) with 
u(11) symmetry \cite{EI1126}, 
the spdf-IBM with u(16) symmetry \cite{EI1126,EI61}, 
and the Vector Boson Model (VBM) with u(6) symmetry \cite{GRR8,GRR9,GRR12}, 
predict in their su(3) limits $\Delta I=1$ staggering of constant 
amplitude, i.e. all the odd levels are raised (or lowered) by the same amount 
of energy with respect to the even levels.  
In other words, $\Delta I=1$ staggering takes alternatively 
positive and negative values of equal absolute value as $I$ increases.  

2) Algebraic models of nuclear structure suitable for the description 
of alpha clustering effects, like the Nuclear Vibron Model (NVM) with 
u(6)$\otimes$u(4) symmetry \cite{DaleyI}, 
also predict in the su(3) limit $\Delta I=1$ staggering of constant
amplitude. 

3) Older experimental work \cite{Schueler,Phillips,Sheline}
on octupole nuclear bands suggests that 
$\Delta I=1$ staggering starts from large values and its amplitude 
decreases with increasing $I$. These findings are in agreement with 
the interpretation that an octupole band is gradually formed as 
angular momentum increases \cite{Leander,LeanderSh}.  

4) Recent work on experimental data for diatomic molecules shows that 
in some rotational bands $\Delta I=1$ staggering of constant amplitude seems 
to appear \cite{RMD}, while in other bands a variety of shapes, 
reminiscent of beats, are exhibited \cite{RMD}. 

Motivated by these recent findings, we make in the present work a systematic 
study in the light actinide region of all octupole bands for which 
at least 12 energy levels are known
\cite{Cocks,Schulz,Ra220,Th222,Th224,Th226,Th228}, 
taking advantage of recent detailed 
experimental work in this region.  
The questions to which we have hoped 
to provide answers are: 

1) Which patters of behaviour of the amplitude of the $\Delta I=1$ 
staggering appear? Are these patterns related to the ones seen in 
diatomic molecules \cite{RMD}? 

2) Can these patterns be interpreted in terms of the existing models
\cite{DaleyI,EI1126,EI61,GRR8,GRR9,GRR12}, 
or in terms of any other theoretical description? 

In Section 2 of the present paper the formalism of staggering is discussed,
and is subsequently applied to the experimental data for octupole bands
of light actinides in Section 3. Section 4 contains the relevant 
predictions of various algebraic models, while an interpretation of the 
experimental observations is given in Section 5. Finally, Section 6 contains 
the conclusions reached, as well as plans for future work. 

{\bf 2. Formalism} 

Traditionally the odd--even staggering ($\Delta I=1$ staggering) in 
octupole bands, as well as in gamma bands, has been estimated 
quantitatively through use of the expression \cite{Nazar}
\begin{equation}
\delta E(I)= E(I)- {(I+1) E(I-1) +I E(I+1)\over 2I+1}, 
\end{equation} % (1) 
where $E(I)$ denotes the energy of the level with angular momentum $I$. 
This expression vanishes for 
\begin{equation}
E(I)=E_0 + A I(I+1),
\end{equation} % (2) 
but not for
\begin{equation} 
E(I)=E_0 + A I(I+1) + B (I(I+1))^2.
\end{equation} % (3) 
 Therefore it is suitable for measuring 
deviations from the pure rotational behaviour. 

Recently, however, a new measure of the magnitude of 
staggering effects has been introduced \cite{Ced}
in the study of $\Delta I=2$ staggering 
of nuclear superdeformed bands. In this case  
the experimentally determined quantities are the
$\gamma$-ray transition energies between levels differing by two units
of angular momentum ($\Delta I=2$). For these the symbol
\begin{equation}  
E_{2,\gamma}(I) = E(I+2)-E(I) 
\end{equation} %\eqno(4)
is used.
The deviation of the $\gamma$-ray transition energies from the
rigid rotator behavior is then measured by the quantity \cite{Ced}
\begin{equation}
 \Delta E_{2,\gamma}(I) = {1\over 16} (6E_{2,\gamma}(I) -4E_{2,\gamma} (I-2)
-4E_{2,\gamma}(I+2) +E_{2,\gamma}(I-4) +E_{2,\gamma}(I+4)). 
\end{equation} %\eqno(5)

\noindent Using the rigid rotator expression of Eq. (2)  one
can easily see that
in this case $\Delta E_{2,\gamma} (I) $ vanishes.
In addition, the perturbed rigid rotator expression of Eq. (3) 
gives vanishing $\Delta E_{2,\gamma} (I)$. 
These properties are due to the fact that Eq. (5) is a (normalized) 
discrete approximation of the fourth derivative of the function
$E_{2,\gamma}(I)$, i.e. essentially the fifth derivative of the 
function $E(I)$. Therefore we conclude that Eq. (5) is a more sensitive 
probe of deviations from rotational behaviour than Eq. (1). 

By analogy, $\Delta I=1$ staggering in nuclei can be measured by the 
quantity 
\begin{equation}
\Delta E_{1,\gamma}(I)= {1\over 16} (6 E_{1,\gamma}(I)-4 E_{1,\gamma}(I-1)
-4 E_{1,\gamma}(I+1) +E_{1,\gamma}(I-2) + E_{1,\gamma}(I+2) ),
\end{equation} %(6)
where 
\begin{equation}
E_{1,\gamma}(I)= E(I+1)-E(I).
\end{equation} %(7)
The transition energies $E_{1,\gamma} (I)$ are determined directly from 
experiment.  

{\bf 3. Analysis of experimental data} 

We have applied the formalism described above to all octupole bands 
of light actinides for which at least 12 energy levels are known
\cite{Cocks,Schulz,Ra220,Th222,Th224,Th226,Th228}
and which show no backbending (i.e. bandcrossing) \cite{deVoigt} behaviour.  
These nuclei are listed in Table 1, along with the relevant values of the 
$R_4$ ratio,
\begin{equation}
R_4 ={E(4)\over E(2)},
\end{equation}
a well known characteristic of collective behaviour. 
Several nuclei ($^{222-226}$Ra, $^{224-228}$Th) are rotational or 
near-rotational (having $10/3 \geq R_4 \geq 2.7$), while others 
($^{218-222}$Rn, $^{220}$Ra, $^{220-222}$Th) 
are vibrational or near-vibrational (having $2.4 \geq R_4 \geq 2$). . 
A special case is $^{218}$Ra, 
for which it has been argued \cite{Schulz} that it is an example of a new
type 
of transitional nuclei, in which the octupole deformation dominates over all 
other types of deformation.

The staggering results for $^{218-222}$Rn, $^{218-226}$Ra, and
$^{220-228}$Th, 
are shown in Fig. 1, Fig. 2, and Fig. 3, respectively. 
In all cases the 
experimental errors are of the size of the symbol used for the experimental
point and therefore are not visible. 
The following observations can be made:

1) In all cases the shapes appearing are consistent with the following 
pattern: $\Delta I=1$ staggering starts from large values at low $I$, 
it gradually decreases down to zero, then it starts increasing again, 
then it decreases down to zero and starts raising again.  
In other words, figures resembling beats appear. The most complete 
``beat'' figures appear in the cases of $^{220}$Ra, $^{224}$Ra, $^{222}$Th, 
as well as in the cases of $^{218}$Ra, $^{222}$Ra, $^{226}$Ra. 

2) In all cases within the first ``beat'' (from the beginning up to 
the first zero of $\Delta E_{1,\gamma}(I)$~) 
the minima appear at odd $I$, 
indicating that in this region the odd levels are slightly raised 
in comparison to the even levels. Within the second ``beat'' (i.e. 
between the first and the second zero of $\Delta E_{1,\gamma}(I)$~),
the opposite holds: the minima appear at even $I$, indicating that 
in this region the odd levels are slightly lowered in comparison to the 
even levels. Within the third ``beat'' (after the second zero of 
$\Delta E_{1,\gamma}(I)$~) the situation occuring within the first 
``beat'' is repeated. (Notice that $^{220}$Th is not an exception, since 
what is seen in the figure is the second ``beat'', starting from $I=6$.) 

3) In the case of $^{222}$Rn the decrease of the staggering with increasing 
$I$, in the region for which experimental data exist, is very slow, giving 
the impression of almost constant staggering. One can get a similar 
impression from parts of the patterns shown, as, for example, in the cases of 
$^{220}$Ra (in the region $I=12-20$), $^{222}$Ra (for $I=9-17$), 
$^{224}$Ra (for $I=10-16$), $^{226}$Ra (for $I=14-20$), $^{222}$Th 
(for $I=10-18$).  

These observations bear considerable similarities to $\Delta I=1$ 
staggering patterns found in rotational bands of diatomic molecules. 
In particular:

1) Staggering patterns of almost constant amplitude have been found in 
some rotational bands of the  AgH \cite{RMD} molecule. 

2) Staggering patterns resembling the ``beat'' structure have been 
seen in several bands of the AgH molecule \cite{RMD}. 

The following comments are also in place: 

1) In all cases bands not influenced by bandcrossing effects \cite{deVoigt}
have been 
considered, in order to make sure that the observed effects are ``pure'' 
single-band effects. 
The only exception is $^{220}$Th, which shows signs of 
bandcrossings at $10^+$ and $13^-$, which, however, do not influence the 
relevant staggering pattern, which is shown in Fig. 3(a) for reasons of 
completeness. A special case is $^{218}$Ra, which shows a rather irregular
dependence of $E(I)$ on $I$. As we have already mentioned, it has been argued 
\cite{Schulz} that this 
nucleus is an example of a new type of transitional nuclei in which 
the octupole deformation dominates over all other types of deformation. 

2) The same ``beat'' pattern appears in both rotational and vibrational 
nuclei. The only slight difference which can be observed, is that the 
first vanishing of the staggering amplitude seems to occur at higher $I$ 
for the rotational isotopes than for their vibrational counterparts. 
Indeed, within the Ra and Th series of isotopes under study, the $I$ at which 
the first vanishing of the staggering amplitude occurs seems to be an 
increasing function of $R_4$, i.e. an increasing function of the quadrupole 
collectivity.  

3) The present findings are partially consistent with older work 
\cite{Schueler,Phillips,Sheline}. 
The limited sets of data of that time were reaching only up to the $I$ at 
which the first vanishing of the staggering amplitude occurs. It was then 
reasonable to assume that the staggering amplitude decreases down to 
zero and remains zero afterwards, since no experimental evidence for 
``beat'' patterns existed at that time.  

{\bf 4. Algebraic models} 

As we have seen in the previous section, certain $\Delta I=1$ staggering 
patterns occur in the octupole bands of the light actinides. Before 
attempting any interpretation of these results, it is instructive to 
examine what kind of staggering patterns are predicted by various 
algebraic models of nuclear structure describing such bands.  
As we have already mentioned, these models are 
related to the description of octupole degrees of freedom, which are 
responsible for the presence of octupole bands, i.e. bands with a 
sequence of levels with $I^{\pi}= 0^+$, $1^-$, $2^+$, $3^-$, $4^+$, $5^-$, 
\dots \cite{Schueler,Phillips,Sheline}.
These bands are thought to be present in cases in which the nucleus 
acquires a shape with octupole deformation, i.e. a pear-like shape
\cite{Leander,LeanderSh}. 

{\it 4.1 The spf-Interacting Boson Model} 

In the spf-IBM \cite{EI1126}, which possesses an u(11) symmetry, 
$s$, $p$, and $f$ bosons (i.e. bosons 
with angular momentum 0, 1, and 3, respectively)
are used. Octupole 
bands are described in the su(3) limit, which corresponds to the chain
\begin{equation}
{\rm u}(11) \supset {\rm u}(10) \supset {\rm su}(3) \supset 
{\rm o}(3) \supset {\rm o}(2).
\end{equation} % (20)
The relevant basis is 
\begin{equation}
\vert N, N_b, \omega_b, (\lambda_b, \mu_b), K_b, I, M >, 
\end{equation} % (21) 
where $N$ is the total number of bosons labelling the irreducible 
representations (irreps) of u(11), 
$N_b$ is the total number of negative parity bosons ($p$ and $f$)
labelling the irreps of u(10), $\omega_b$ is the ``missing'' quantum 
number in the decomposition u(10)$\supset$su(3), 
$(\lambda_b, \mu_b)$ are the Elliott
quantum numbers \cite{Elliott}
labelling the irreps of su(3), $K_b$ is the ``missing'' 
quantum number in the decomposition su(3)$\supset$o(3) \cite{Elliott}, 
$I$ is the angular 
momentum quantum number labelling the irreps of o(3), $M$ is the 
$z$-component of the angular momentum labelling the irreps of o(2). 
The energy eigenvalues are given by
\begin{equation}
 E(N_b, \lambda_b, \mu_b, I)= \alpha +\beta N_b +\gamma N_b^2 
+\kappa C(\lambda_b,\mu_b)+\kappa' I(I+1),
\end{equation} % (22) 
where
\begin{equation}
C(\lambda,\mu)= \lambda^2 +\mu^2+ \lambda \mu+3\lambda +3\mu.
\end{equation} % (23) 

It is clear that positive parity states occur when $N_b$ is even, while 
negative parity states occur when $N_b$ is odd. In the case of $N$ being
even, the ground state band is sitting in the $(3N,0)$ irrep, while 
the odd levels of negative parity are sitting in the $(3N-3,0)$ irrep. 
Then from Eq. (6) one obtains
\begin{equation}
\Delta E(I)=\cases{ -(\beta+\gamma (2N-1)+18\kappa N), & for $I=$ even, \cr
                      +(\beta+\gamma (2N-1)+18\kappa N), & for $I=$ odd. \cr}
\end{equation} % (24) 
In the case of $N$ being odd, the ground state band is sitting in the 
$(3N-3,0)$ irrep, while the odd levels of negative parity are sitting in the 
$(3N,0)$ irrep. Then from Eq. (6) one has 
\begin{equation}
\Delta E(I)=\cases{ +(\beta+\gamma (2N-1)+18\kappa N), & for $I=$ even, \cr
                       -(\beta+\gamma (2N-1)+18\kappa N), & for $I=$ odd. \cr}
\end{equation} % (25) 
Since $N$ is a constant for a given nucleus, expressing the number of valence 
nucleon pairs counted from the nearest closed shells \cite{IA}, 
we see that $\Delta I=1$ 
staggering of constant amplitude is predicted. 

{\it 4.2 The spdf-Interacting Boson Model}

In the spdf-Interacting Boson Model \cite{EI1126,EI61}, which possesses a
u(16) symmetry, $s$, $p$, $d$, and $f$ bosons (i.e. 
bosons with angular momentum 0, 1, 2, and 3, respectively) are taken into 
account. Octupole bands are described in the su(3) limit, which corresponds 
to the chain
\begin{equation}
{\rm u}(16) \supset {\rm u_a}(6) \otimes {\rm u_b}(10) \supset 
{\rm su_a}(3) \otimes {\rm su_b}(3)
\supset {\rm su}(3) \supset {\rm o}(3) \supset {\rm o}(2).
\end{equation} % (26)
The relevant basis is
\begin{equation}
\vert N, N_a, N_b, \omega_b, (\lambda_a, \mu_a), (\lambda_b, \mu_b), 
(\lambda,\mu), K, I, M>,
\end{equation} % (27) 
where $N$ is the total number of bosons labelling the irreps of u(16), 
$N_a$ is the number of positive parity bosons labelling the irreps of
u$_a$(6),
and $N_b$ is the number of negative parity bosons labelling the irreps of 
u$_b$(10). The rest of the quantum numbers are analogous to those appearing 
in the basis of the u(11) model, described above. su(3) is the algebra
obtained by adding the corresponding generators of su$_a$(3) and su$_b$(3).  
The energy eigenvalues are given by
$$E(N_b, \lambda_a, \mu_a, \lambda_b, \mu_b, \lambda, \mu, I)=$$
\begin{equation}
\alpha +\beta N_b +\gamma N_b^2+ \kappa_a C(\lambda_a,\mu_a) 
+\kappa_b C(\lambda_b,\mu_b)+\kappa C(\lambda,\mu)+\kappa' I(I+1),
\end{equation} % (28)
with $C(\lambda,\mu)$ defined as in Eq. (12). 

The ground state band is sitting in the $(2N,0)_a$ irrep (which contains 
$N$ bosons of positive parity and no bosons of negative parity), 
while the odd levels of negative parity are sitting in the 
$(2N-2,0)_a$ $(3,0)_b$ $(2N+1,0)$ band (which contains $N-1$ bosons of
positive parity and one boson of negative parity). 
Then from Eq. (6) one has
\begin{equation}
\Delta E(I)= \cases{ +(\beta +\gamma-2 k_a (4N+1)+ 18 k_b+4 k (N+1)) &
for $I=$ even,\cr         -(\beta +\gamma-2 k_a (4N+1)+ 18 k_b+4 k (N+1)) &
for $I=$ odd.\cr}
\end{equation} % (29) 
Therefore $\Delta I=1$ staggering of constant amplitude is predicted, 
since $N$ is a constant for a given nucleus, representing the number of
valence nucleon pairs counted from the nearest closed shells \cite{IA}. 

Another limit of the spdf-IBM in which octupole bands occur is the o(4) 
limit \cite{EI61}, which corresponds to the chain 
\begin{equation}
{\rm u}(16) \supset {\rm u}(4)_a \otimes {\rm u}(4)_b \supset 
{\rm sp}(4)_a \otimes {\rm sp}(4)_b \supset {\rm su}(2)_a \otimes 
{\rm su}(2)_b \supset {\rm o}(3) \supset {\rm o}(2), 
\end{equation}
and owes its name to the isomorphism 
\begin{equation}
{\rm su}(2)_a \otimes {\rm su}(2)_b \approx {\rm o}(4).
\end{equation}
The relevant basis is 
\begin{equation}
| N, (n_1, n_2, n_3, n_4), (n'_{1a}, n'_{2a}), (n'_{1b}, n'_{2b}), 
\nu, j_a, j_b, I, M>,
\end{equation}
where $N$ is the total number of bosons labelling the irreps of u(16), 
$(n_1, n_2, n_3, n_4)$ are labelling the irreps of u(4)$_a$ and u(4)$_b$, 
$(n'_{1a}, n'_{2a})$ and $(n'_{1b}, n'_{2b})$ are labelling the irreps of 
sp(4)$_a$ and sp(4)$_b$ respectively, $\nu$ denotes the three missing 
quantum numbers required in this case, $j_a$ and $j_b$ label the irreps of
su(2)$_a$ and su(2)$_b$ respectively, while $I$ and $M$  have the same 
meaning as before. The energy eigenvalues are given by 
$$ E(N, n_1, n_2, n_3, n_4, n'_{1a}, n'_{2a}, n'_{1b}, n'_{2b}, \nu, 
j_a, j_b, I, M) $$
\begin{equation}
= E_0-2A [j_a (j_a+1) +j_b (j_b+1)] + (B+A) I(I+1)
= E_0 -A [\omega (\omega+2)+(\omega')^2]+ (B+A) I(I+1), 
\end{equation}
where $(\omega, \omega')$ are labelling the irreps of o(4) and are 
connected to $j_a$ and $j_b$ through the relations 
\begin{equation}
\omega = j_a +j_b, \qquad \omega' =|j_a-j_b|.
\end{equation}
The lowest lying irrep is the irrep $(3N,0)$, which contains states 
of positive parity and states of negative parity together, i.e. it contains 
the states $0^+$, $1^-$, $2^+$, $3^-$, $4^+$, $5^-$, \dots, up to the state 
with $I=3N$. It is clear that in this case Eq. (6) gives a vanishing result, 
i.e. no $\Delta I=1$ staggering occurs in this limit. 

{\it 4.3 The Vector Boson Model} 

In the Vector Boson Model (VBM) \cite{GRR8,GRR9,GRR12}, 
the collective states are 
described in terms of two distinct kinds of vector bosons, whose creation 
operators $\mbox{\boldmath $\xi^+$ }$ and $\mbox {\boldmath $\eta^+$ }$ 
are o(3) vectors and in addition transform according to two independent 
su(3) irreducible representations (irreps) of the type $(\lambda, \mu)
=(1,0)$, i.e. they are two distinct bosons of angular momentum 1.  
Octupole bands are described 
in the su(3) limit of the VBM, which corresponds to the chain 
\begin{equation}
{\rm u}(6) \supset {\rm su}(3)  \otimes {\rm u}(2) \supset 
{\rm so}(3) \otimes {\rm u}(1).
\end{equation} % (30) 
The relevant basis is 
\begin{equation} 
\vert N, (\lambda,\mu), (N,T), K, I, T_0 >,
\end{equation} % (31) 
where $N$ is the total number of bosons labelling the irreps 
of u(6), $(\lambda,\mu)$ are the Elliott quantum numbers \cite{Elliott}
labelling the irreps of su(3), $N$ and $T$ are the quantum numbers labelling 
the irreps of u(2), $K$ is the ``missing'' quantum number in the 
su(3)$\supset$so(3) decomposition \cite{Elliott}, $I$ is the angular momentum 
quantum number labelling the irreps of so(3), and $T_0$ is the 
pseudospin projection quantum number labelling the irreps of u(1). 
The algebras su(3) and u(2) are mutually complementary 
\cite{Quesne1,Quesne2,Quesne3}, 
their irreps $(\lambda,\mu)$ and $(N,T)$ being related by 
\begin{equation}
N=\lambda+2\mu, \qquad  T=\lambda/2.
\end{equation} % (32)  
The energy eigenvalues are given by 
\begin{equation} 
E(N, \lambda, \mu, K, I, T_0=T)= a N+ a_6 N(N+5)
+ a_3  C(\lambda,\mu) + b_3 I(I+1) +a_1 {\lambda^2\over 4},
\end{equation} % (33) 
with $C(\lambda,\mu)$ defined as in Eq. (12). 

The ground state band is sitting in the $(0,\mu)=\left(0,{N\over2}\right)$ 
irrep of su(3), while the odd levels of negative parity are sitting in the 
$(2,\mu-1)=\left(2,{N\over 2}-1\right)$ irrep.  
Then from Eq. (6)  one obtains
\begin{equation}  
\Delta E(I)=\cases{ +(6 a_3+a_1),&  for $I=$ even, \cr
                        -(6 a_3+a_1), & for $I=$ odd.\cr}
\end{equation} % (34) 
Therefore $\Delta I=1$ staggering of constant amplitude is predicted. 

{\it 4.4 The Nuclear Vibron Model} 

As we have already mentioned, an alternative interpretation of the low lying 
negative parity states appearing in the light actinides has been given 
following the assumption that alpha clustering is important in this region
\cite{DaleyI,Buck}. 
An algebraic model appropriate for the description of clustering effects 
in nuclei is the Nuclear Vibron Model \cite{DaleyI}, 
which uses $s$ and $d$ bosons 
for the description of nuclear collectivity, plus $s'$ and $p$ bosons 
for taking into account the distance separating the center of the cluster 
from the center of the remaining nucleus. The chain corresponding to the 
su(3) limit of this model is
\begin{equation}
{\rm u}(6) \otimes {\rm u}(4) \supset {\rm su_a}(3)\otimes {\rm u_b}(3) 
\supset {\rm su_a}(3) \otimes {\rm su_b}(3) \supset {\rm su}(3) \supset
{\rm o}(3) \supset {\rm o}(2),
\end{equation} 
where the subscript {\rm a} labels the subalgebras of u(6), while 
the subscript {\rm b} labels the subalgebras of u(4). 
The relevant basis is 
\begin{equation}
| N, M, (\lambda_a,\mu_a), n_p, (\lambda,\mu), \chi, I, M>,
\end{equation} 
where $N$ is the number of the $s$ and $d$ bosons related to the u(6) 
algebra, $M$ is the number of the $s'$ and $p$ bosons related to the u(4) 
algebra, $(\lambda_a, \mu_a)$ are the Elliott quantum numbers 
\cite{Elliott} related to 
su$_a$(3), $n_p$ is the number of $p$ bosons, $(\lambda, \mu)$ are the 
Elliott quantum numbers related to su(3), $\chi$ is the Vergados ``missing''
quantum number \cite{Vergados}
in the decomposition su(3)$\supset$o(3), while $I$ and $M$ 
represent the angular momentum and its $z$-component respectively, as usual. 
The energy eigenvalues are given by
\begin{equation}
E(n_p, \lambda_a, \mu_a, \lambda, \mu, I)=
\epsilon_p n_p + \alpha_p n_p (n_p+3) +\kappa_d C(\lambda_a, \mu_a)
+\kappa C(\lambda, \mu) + \kappa' I(I+1),
\end{equation}
with $C(\lambda,\mu)$ defined as in Eq. (12). 

The ground state band is characterized by $(\lambda_a, \mu_a)= 
(2N,0)$, $n_p=0$, $(\lambda, \mu)=(2N,0)$ (i.e. it contains $N$ bosons 
of positive parity and no $p$-boson of negative parity), 
while the negative parity band 
is characterized by $(\lambda_a, \mu_a) = (2N,0)$, $n_p=1$, 
$(\lambda, \mu)=(2N+1,0)$ (i.e. it contains $N$ bosons of positive parity 
plus one $p$-boson of negative parity). 
Then from Eq. (6) one has 
\begin{equation} 
\Delta E(I)= 
\cases{+(\epsilon_p+4 \alpha_p+4 \kappa (N+1)), &  for $I=$ even, \cr
       -(\epsilon_p+4 \alpha_p+4 \kappa (N+1)),  &  for $I=$ odd.    \cr}
\end{equation} 
Therefore $\Delta I=1$ staggering of constant amplitude is predicted. 

{\it 4.5 Discussion}

We conclude that the various algebraic models, 
describing low lying negative parity bands in terms of octupole 
deformation \cite{EI1126,EI61,GRR8,GRR9,GRR12}
or in terms of alpha clustering \cite{DaleyI}, 
predict in their su(3) limits odd--even staggering 
($\Delta I=1$ staggering) of constant amplitude. In all cases the 
staggering results from the fact that the negative parity states belong 
to an irrep different from the one in which the positive parity states 
composing the ground state band sit. 

It should be noticed, as already remarked in Section 3, that the experimental 
data indicate that the value of $I$
at which the first vanishing of the staggering amplitude occurs increases 
as a function of $R_4$, i.e. as the rotational limit is approached. 
The higher the value of $I$ at which the first vanishing occurs, 
the more smooth the decrease of the staggering as a function of $I$ is.
We see, therefore, that as the rotational limit is approached, the 
experimental data approach more and more the constant staggering 
prediction provided by the various algebraic models. The best 
example is provided by $^{228}$Th, the most rotational among the nuclei 
studied here.  

As far as limits of algebraic models different from the su(3) limit are 
concerned, no staggering occurs in the o(4) limit of the spdf-IBM, which 
has been fully worked out \cite{EI61}. 
Working out the details of other non-su(3) limits, 
like the ones of the Vector Boson Model mentioned in Ref. \cite{GRR8}, 
is an interesting open problem. 

{\bf 5. Interpretation of the experimental observations} 

Although the results of the  previous section 
are sufficient for providing an explanation for 
$\Delta I=1$ staggering in the cases in which this appears as having 
almost constant amplitude, it is clear that some additional thinking 
is required for the many cases in which the experimental results show 
a ``beat'' pattern, as in Section 3 has been exhibited. 

A simple explanation for the appearance of ``beat'' patterns can be given 
by the following assumptions:

1) It is clear that in each nucleus the even levels form the ground state
band,
which starts at zero energy, 
while the odd levels form a separate negative parity band, which starts at 
some higher energy. Let us call $E_0$ the bandhead energy of the negative 
parity band. 

2) It is reasonable to try to describe the ground state band by an expression 
like 
\begin{equation}
E_+(I)=A I(I+1) -B (I(I+1))^2 + C (I(I+1))^3 + \cdots
\end{equation}
where the subscript $+$ reminds us of the positive parity of these levels. 
Such expansions in terms of powers of $I(I+1)$ have been long used for the 
description of nuclear collective bands \cite{Xu}. 
They also occur if one considers \cite{PLB251}
Taylor expansions of the energy expressions provided by the Variable 
Moment of Inertia (VMI) model \cite{VMI} and the su$_q$(2) model
\cite{RRS}. 
Notice that fits to experimental data \cite{Xu} indicate that one always has 
$A > 0$, $B>0$, $C>0$, \dots, while $A$ is usually 3 orders of 
magnitude larger than $B$, $B$ is 3 orders of magnitude larger 
than $C$, etc. 
Eq. (33) has been long used in molecular spectroscopy as well, under the 
name of Dunham expansion \cite{Dunham}. 

3) In a similar way, it is reasonable to try to describe the negative parity 
levels by an expression like 
\begin{equation}
E_-(I)= E_0+ A' I(I+1) - B' (I(I+1))^2 + C' (I(I+1))^3 +\cdots
\end{equation} 
where the subscript $-$ reminds us of the negative parity of these levels, 
while $E_0$ is the above mentioned bandhead energy. 
In analogy to the previous case one expects to have $A'>0$, $B'>0$, $C'>0$, 
\dots 

4) In the above expansions it is reasonable to assume that 
$A>A'$, $B>B'$, $C>C'$, \dots. 
This assumption is in agreement with earlier work \cite{Neer1,Neer2,Vogel}, 
in which the Coriolis couplings between the lowest $K=0$ negative parity band 
and higher negative parity bands with $K\neq 0$ are taken into account, 
resulting in an increase of the monent of inertia of the lowest $K=0$ negative
parity band \cite{RohozG}. This argument means that the coefficient $A'$ 
in Eq. (34), which is inversely proportional to the moment of inertia 
of the negative parity band, should be smaller than the coefficient $A$ 
in Eq. (33), which is inversely proportional to the moment of inertia 
of the positive parity band. In analogy to the relation $A>A'$, which we 
just justified, one can assume $B>B'$, $C>C'$, \dots. This last argument 
is admitedly a weak one, which is however driving to interesting results, 
as we shall soon see. 

Using Eqs (33) and (34) in Eqs (6) and (7) we find the following results 
$$
\Delta E(I)= E_0 -(A-A') (I^2+2I+2) + (B-B') \left(I^4+4 I^3 +13 I^2 +18 I + 
{23\over 2}\right)$$ 
$$
-(C-C') \left( I^6+ 6 I^5 + 33 I^4 + 92 I^3 + 
{357\over 2} I^2 + {333\over 2} I + 68\right) $$
\begin{equation}
+ 45 C' (I+1) +\cdots,  
\quad {\rm for} \quad I={\rm even},
\end{equation}
$$
\Delta E(I)= -E_0 +(A-A') (I^2 + 2 I +2) -(B-B')\left( I^4+ 4 I^3+13 I^2+18 I+
{23\over 2}\right)$$  
$$
+(C-C') \left( I^6+6 I^5+33 I^4+ 92 I^3 + {357\over 2}
I^2 + {333\over 2} I +68\right) $$
\begin{equation}
- 45 C' (I+1) +\cdots,
\quad {\rm for} \quad I={\rm odd}.
\end{equation}
A sample staggering pattern drawn using these formulae is shown in Fig. 4. 
On these results the following comments can be made:

1) The expression for odd $I$ is the opposite of the expression with 
even $I$. This explains why in Fig. 4 the staggering points for even $I$
and the staggering points for odd $I$ form two lines which are reflection 
symmetric with respect to the horizontal axis. 

2) For even $I$ the behaviour of the staggering amplitude is as follows:
At low $I$ it starts from a positive value, because of the presence of 
$E_0$. As $I$ increases, the second term, which is essentialy proportional 
to $I^2$, becomes important. ($E_0$ is expected to be much larger than 
$(A-A')$.)
This term is negative (since $A>A'$), thus 
it decreases the amplitude down to negative values. At higher values 
of $I$ the third term, which is essentially proportional to $I^4$, 
becomes important. (Remember that usually $B$ is 3 orders 
of magnitude smaller than $A$ \cite{Xu}.)
This term is positive (since $B>B'$), thus it 
increases the amplitude up to positive values. 
(The behaviour up to this point can be seen in Fig. 4.)
At even higher 
values of $I$ the fourth term, which is essentially proportional 
to $I^6$, becomes important. (Remember that usually $C$ is 3 orders of 
magnitude smaller than $B$ \cite{Xu}.) 
This term is negative (since $C>C'$), thus 
it decreases the amplitude again down to negative values, and so on. 

3) For odd $I$ the behaviour of the staggering amplitude is exactly the 
opposite
of the one described in 2) for even $I$. The amplitude starts from a negative 
value and then becomes consequently positive (because of the second term), 
negative (because of the third term), again positive (because of the 
fourth term), and so on. The first three steps of this behaviour can be seen 
in Fig. 4.

4) When drawing the staggering figure one jumps from an even $I$ to an 
odd $I$, then back to an even $I$, then back to an odd $I$, and so on. 
It is clear therefore that a ``beat'' pattern appears, as it is seen in 
Fig. 4. 

The following additional comments are also in place: 

1) In the case of a single band (i.e. in the case of $A=A'$, $B=B'$, 
$C=C'$, etc), the first contribution to the staggering measure 
$\Delta E(I)$ is the last term in Eqs (35), (36), which comes from the 
$C (I(I+1))^3$ term in the energy expansion (see Eqs (33), (34)~). 
This is understandable: Since Eq. (6) is a discrete approximation of the 
fifth derivative of the function $E(I)$, as it has already been remarked, 
the terms up to $B(I(I+1))^2$ are ``killed'' by the derivative, 
while the $C (I(I+1))^3$ term gives a contribution linear in $I$. 

2) The last term in Eqs (35), (36)
does not influence significantly the behaviour 
of the staggering pattern, since $C$ is usually 6 orders of magnitude 
smaller than $A$ and 3 orders of magnitude smaller than $B$ \cite{Xu}. 

3) One could argue that the above reasoning is valid only for the case 
of rotational or near-rotational bands, for which the expansions 
of Eqs (33), (34) are known to be adequate (although one should be reminded 
at this point that the VMI model describes quite well not only 
rotational, but also transitional and even vibrational nuclei). 
One can attempt to mend this problem by adding to the expansions of 
Eqs (33) and (34) a linear term, in the spirit of the Ejiri formula
\cite{Ejiri}, 
the Variable Anharmonic Vibrator Model (VAVM) \cite{VAVM}, 
and the u(5) and o(6) limits of the Interacting Boson Model \cite{PRC50}
\begin{equation}
E_+(I)= A_1 I + A I(I+1)- B (I(I+1))^2 +C (I(I+1))^3 + \cdots,
\end{equation}
\begin{equation}
E_-(I)= E_0 +A'_1 I + A' I(I+1) - B' (I(I+1))^2 +C' (I(I+1))^3 + \cdots. 
\end{equation} 
Then Eqs (35) and (36) get modified as follows
$$
\Delta E(I)= E_0 -(A_1-A'_1) \left(I+{1\over 2}\right) - (A-A')
(I^2+2I+2) $$
\begin{equation}
+ (B-B') \left( I^4+4I^3+13I^2+18I+{23\over 2}\right)-\cdots,
\quad {\rm for} \quad I={\rm even},
\end{equation}
$$
\Delta E(I)= -E_0 +(A_1-A'_1) \left( I+{1\over 2}\right) + (A-A') (I^2+2I+2)$$
\begin{equation}
-(B-B') \left( I^4+4I^3+13I^2+18I+{23\over 2}\right)+\cdots,
\quad {\rm for} \quad I={\rm odd}.
\end{equation} 
We see that the extra term, which is proportional to $(A_1-A'_1)$, 
plays the same role as the term proportional to $(A-A')$ in shaping 
up the behaviour of the staggering amplitude. Therefore the conclusions 
reached above for rotational nuclei apply equally well to vibrational and 
transitional nuclei as well.

4) This type of explanation of the staggering patterns seems to be 
outside the realm of the form of the su(3) limits of the algebraic models 
presented above. Even if one decides to include 
higher order terms of the type $(I(I+1))^2$, $(I(I+1))^3$, etc, in these 
models, by including in the Hamiltonian higher powers of the relevant Casimir 
operator, these terms will appear with the same coefficients for both 
the ground state band and the negative parity band, even though these 
two bands belong to different irreps.  The only possible contributions 
to the staggering will then come from terms like the last term in Eqs (35)
and (36),
which comes from the term $(I(I+1))^3$, and similar terms coming from 
higher powers of $I(I+1)$. However, the term $(I(I+1))^3$ in the framework 
of the algebraic models already corresponds to 6-body interactions \cite{IA}, 
which are usually avoided in nuclear structure studies. 

We conclude therefore that the ``beat'' pattern can be explained in terms 
of two Dunham expansions with slightly different sets of coefficients, 
one for the ground state band with quadrupole deformation and another 
for the negative parity band in which in addition the octupole deformation 
appears. This is, however, a phenomenological finding, the microscopic 
origins of which should be searched for. On this open problem the 
following comments apply:

1) As it has been mentioned above, the Coriolis coupling between the lowest 
$K=0$ negative parity band and higher $K\neq 0$ negative parity bands
\cite{Neer1,Neer2,Vogel} results in an increase of the moment of inertia 
of the lowest $K=0$ negative parity band \cite{RohozG}, 
offering in this way an argument in favor of using 
different coefficients in the Dunham expansions for the negative parity 
states and the positive parity states of the octupole band. However, this 
argument holds for the coefficients of the $I(I+1)$ terms only. If 
Coriolis coupling leads to different coefficients for the rest of the terms 
of the Dunham expansion remains to be seen. 

2) Nuclei with octupole deformation (pear-shaped nuclei) are supposed to be 
described by double well potentials, the relative displacement of the negative
parity levels and the positive parity levels being attributed 
to the tunneling through the barrier separating the wells 
\cite{Leander,LeanderSh,Krappe}. The relative displacement vanishes in the 
limit in which the barrier separating the two wells becomes infinitely high.
It should be examined if the details of the relevant potentials 
\cite{Leander,LeanderSh,Krappe}
give rise to a ``beating'' behavior of the relevant displacement. 

3) The coupling between the quadrupole modes and the octupole modes 
can also give rise to relative displacement of the negative parity levels 
and the positive parity levels of the octupole band \cite{Gajda}.
In this case the octupole deformation can be parametrized in the way 
described in Refs \cite{Ham1,Ham2,Rohoz}. It should be examined if ``beating''
patterns appear in this case. Work in this direction is in progress. 

{\bf 6. Discussion} 

We have demonstrated that octupole bands in the light actinides 
exhibit $\Delta I=1$ staggering (odd--even staggering), 
the amplitude of which shows a ``beat''
behaviour. The same pattern appears in both vibrational and rotational 
nuclei, forcing us to modify the traditional belief that in octupole bands
the 
staggering pattern 
is gradually falling down to zero as a function of the angular momentum $I$ 
and then remains there. 

It has also been demonstrated that the su(3) limits of 
various algebraic models, including 
octupole degrees of freedom \cite{EI1126,EI61,GRR8,GRR9,GRR12}
or based on the assumption that alpha 
clustering is important in this region \cite{DaleyI}, 
predict $\Delta I=1$ staggering 
of amplitude constant as a function of the angular momentum $I$. 
Although this description becomes reasonable in the rotational limit, 
it cannot explain the ``beat'' patterns appearing in both the rotational 
and the vibrational regions. The detailed study of limits other than 
the su(3) ones for these models remains an interesting open problem. 

A simple explanatation of the ``beat'' behaviour has been given by 
describing the even $I$ levels of the ground state band and the odd $I$
levels of the negative parity band by two Dunham expansions \cite{Dunham}
(expansions in powers of $I(I+1)$) with slightly different sets of 
coefficients, the difference 
in the coefficients being attributed to Coriolis couplings of the negative 
parity band to other negative parity bands. However, the microscopic origins
of the ``beat'' behavior need further elucidation, for example in the ways 
mentioned at the end of Section 5. 

The ``beat'' patterns found here in the octupole bands of the light actinides 
bear striking similarities to the ``beat'' patterns seen in the rotational
bands of some diatomic molecules, like AgH \cite{RMD}. It is expected that an 
explanation of the ``beat''
behaviour in terms of two Dunham expansions with slightly different sets of 
coefficients should be equally applicable in this case. 

It is also of interest to check if ``beat'' patterns appear in other kinds of 
bands as well. Preliminary results indicate that such patterns appear in some 
gamma bands ($^{164}$Er, $^{170}$Yb), as well as in a variety of 
negative parity bands. Further work in this direction is needed. 

{\bf Acknowledgements} 

One of the authors (PPR) acknowledges support from the Bulgarian Ministry 
of Science and Education under contract $\Phi$-547.   
Another author (NM) has been supported by the Bulgarian National Fund 
for Scientific Research under contract no MU-F-02/98. 
Three authors (DB,CD,NK) have been supported by the Greek Secretariat 
of Research and Technology under contract PENED 95/1981.

\newpage
%%%%%%%%%%%%%%%%%%%%%%%%%%%%%%%%%%%%%%%%%%%%%%%%%%%%%%%%%%%%%%%%%%%%%%
%%%%%%%%%%%%%%%%%%% Table 1  %%%%%%%%%%%%%%%%%%%%%%%%%%%%%%%%%%%%%%%%

\begin{table}

\caption{ Nuclei included in the study and their $R_4=E(4)/E(2)$ ratios 
(Eq. (8)~).} 
\bigskip

\centering
\begin{tabular}{ | c c | c c | c c | }
\hline
nucleus     & $R_4$ & nucleus    & $R_4$ & nucleus    & $R_4$ \\
\hline
$^{218}$Rn  & 2.014 & $^{218}$Ra & 1.905 & $^{220}$Th & 2.035 \\ 
$^{220}$Rn  & 2.214 & $^{220}$Ra & 2.298 & $^{222}$Th & 2.399 \\ 
$^{222}$Rn  & 2.408 & $^{222}$Ra & 2.715 & $^{224}$Th & 2.896 \\ 
            &       & $^{224}$Ra & 2.970 & $^{226}$Th & 3.136 \\  
            &       & $^{226}$Ra & 3.127 & $^{228}$Th & 3.235 \\  
\hline
\end{tabular}
\end{table}

\bigskip\bigskip

\centerline{\bf Figure captions}

\begin{itemize}

\item[{\bf Fig. 1}] $\Delta E_1(I)$ (in keV), calculated from Eq. (6),
for octupole bands of a) $^{218}$Rn \cite{Cocks}, b) $^{220}$Rn \cite{Cocks}, 
and c) $^{222}$Rn \cite{Cocks}.
The experimental error in all cases is of the order of the symbol used for 
the experimental point and therefore is not seen. See Section 3 for 
discussion. 

\item[{\bf Fig. 2}] Same as Fig. 1, but for a) $^{218}$Ra \cite{Schulz}, b) 
$^{220}$Ra \cite{Ra220}, c) $^{222}$Ra \cite{Cocks}, d) $^{224}$Ra 
\cite{Cocks}, and e) $^{226}$Ra \cite{Cocks}. 

\item[{\bf Fig. 3}] Same as Fig. 1, but for a) $^{220}$Th \cite{Ra220}, 
b) $^{222}$Th \cite{Th222}, c) $^{224}$Th \cite{Th224}, d) $^{226}$Th 
\cite{Th226}, and e) $^{228}$Th \cite{Th228}. 

\item[{\bf Fig. 4}] $\Delta E_1(I)$, calculated from Eq. (6), using 
for the levels with even $I$ the expansion of Eq. (33) with 
$A=10$, $B=5\ 10^{-4}$, $C=0$, and for the levels with odd $I$ the 
expansion of Eq. (34) with $E_0=200$, $A'=9$, $B'=10^{-4}$, $C'=0$. 
See Section 5 for discussion. 

\end{itemize}

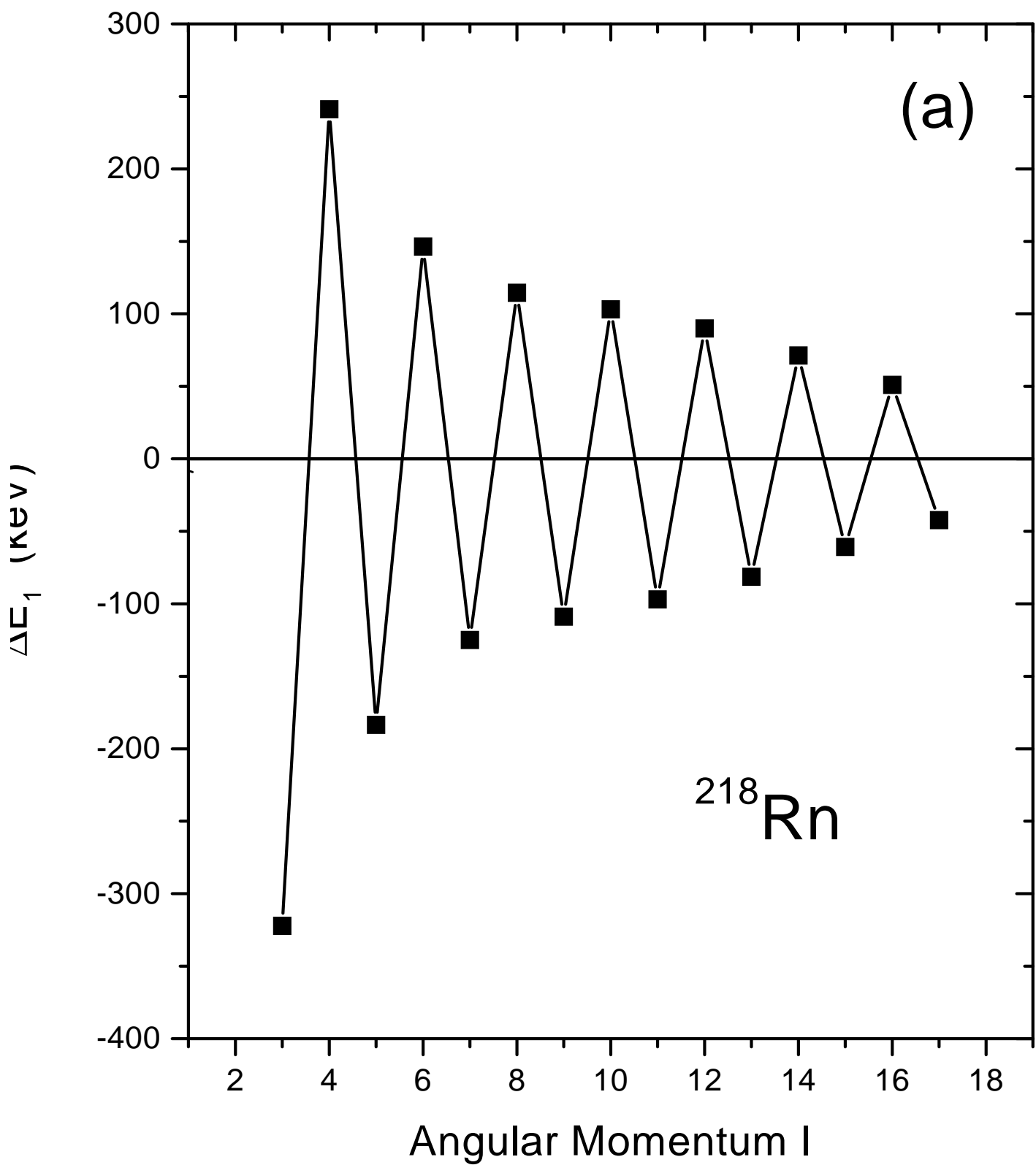
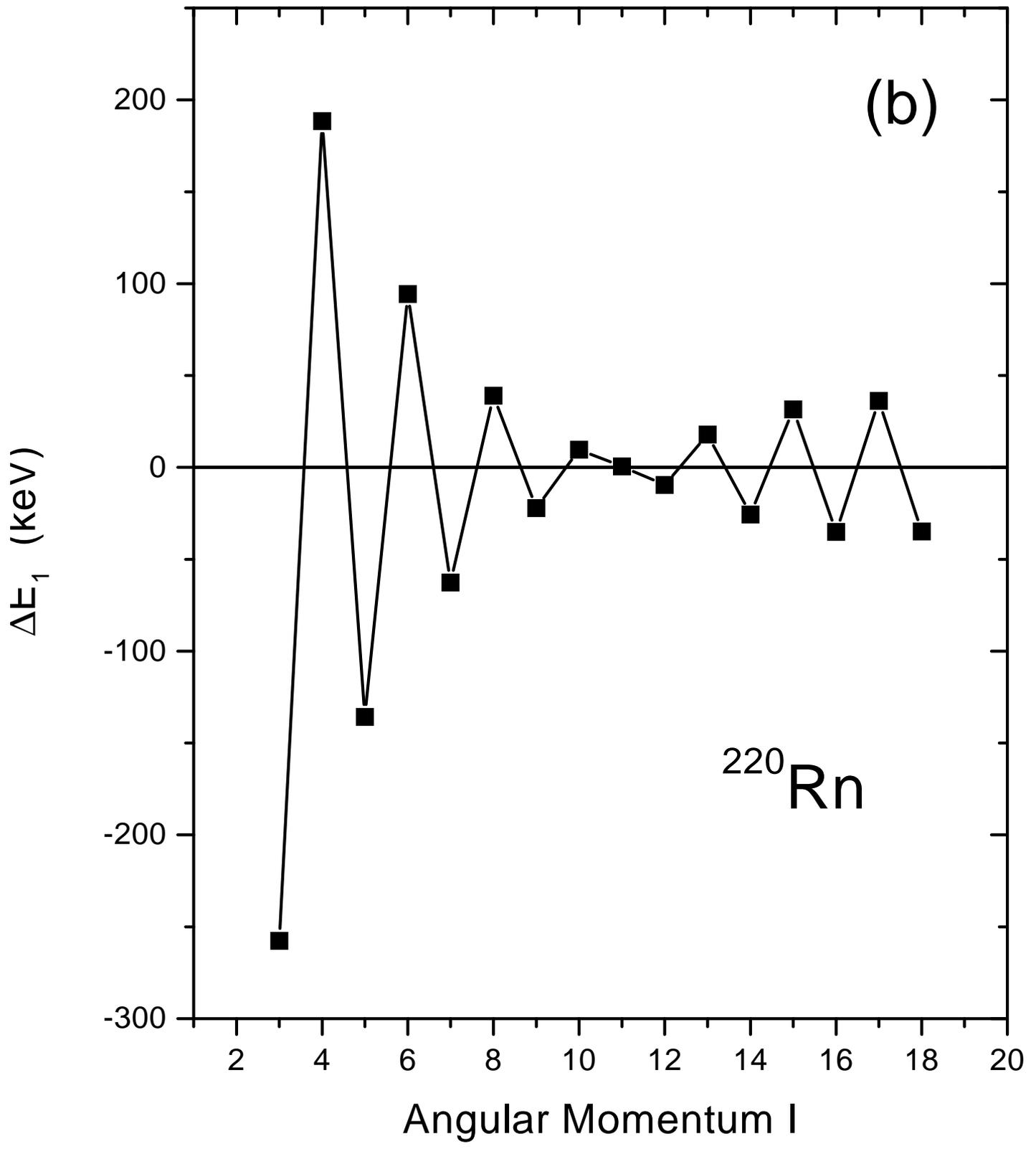
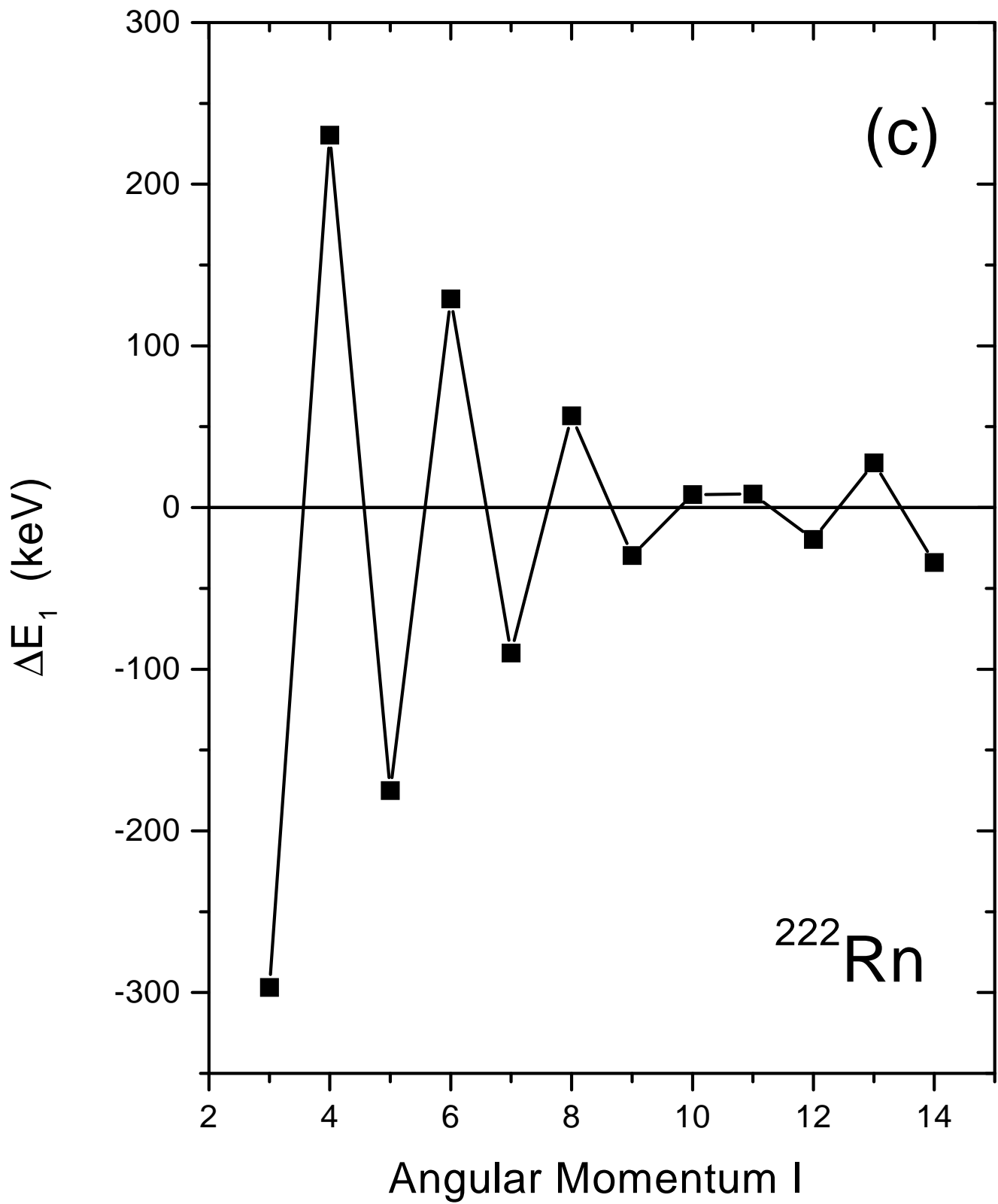
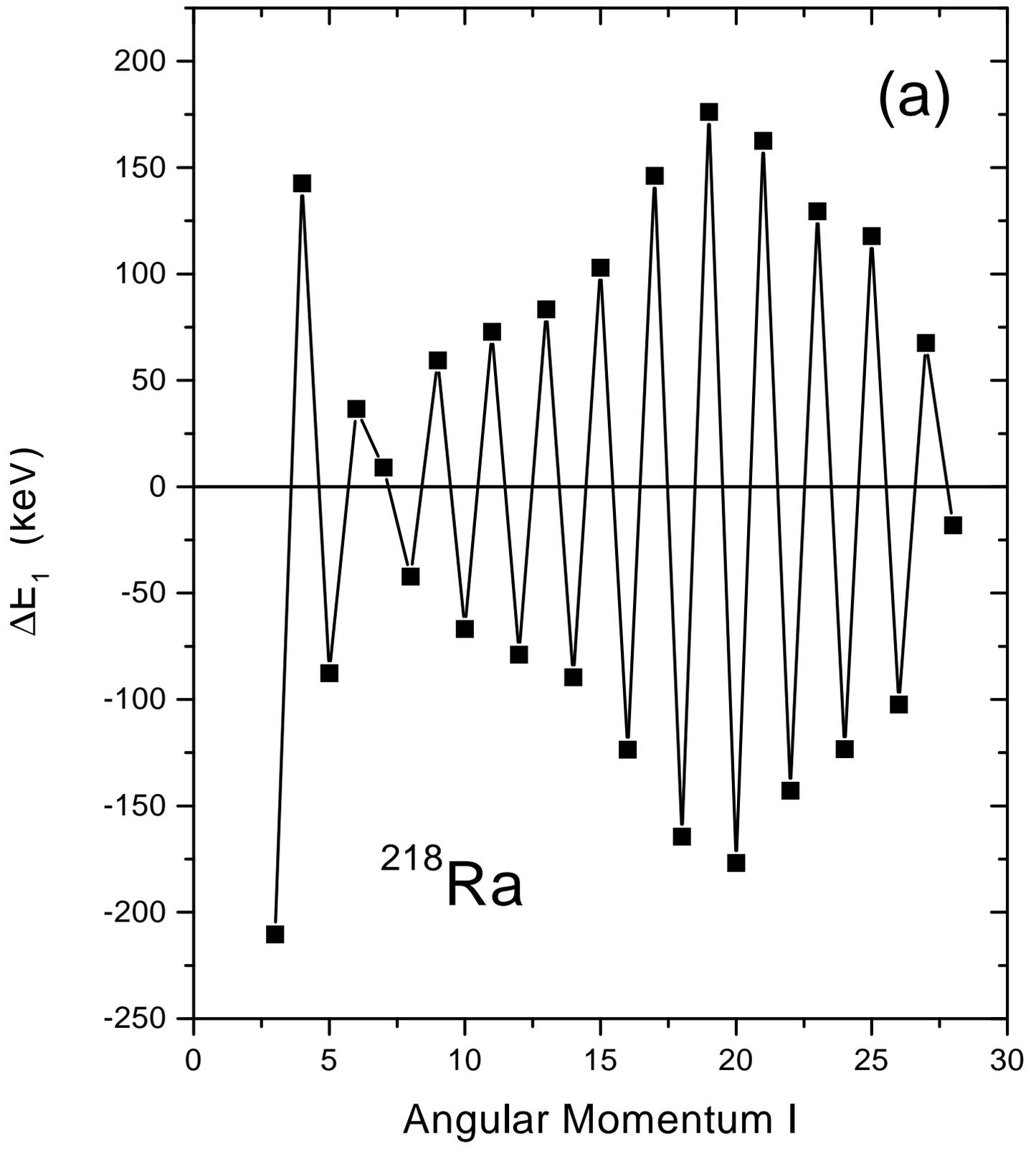
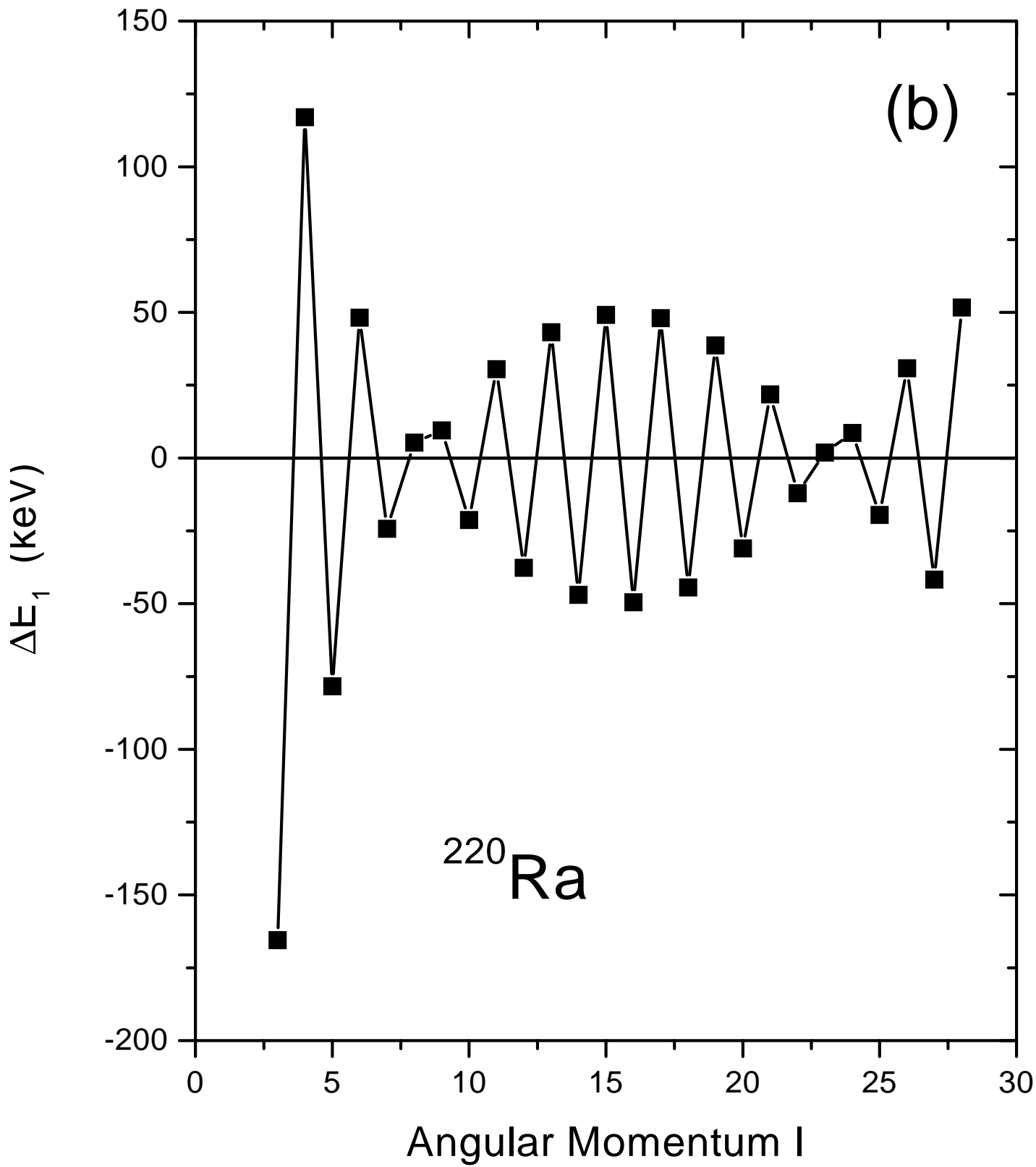
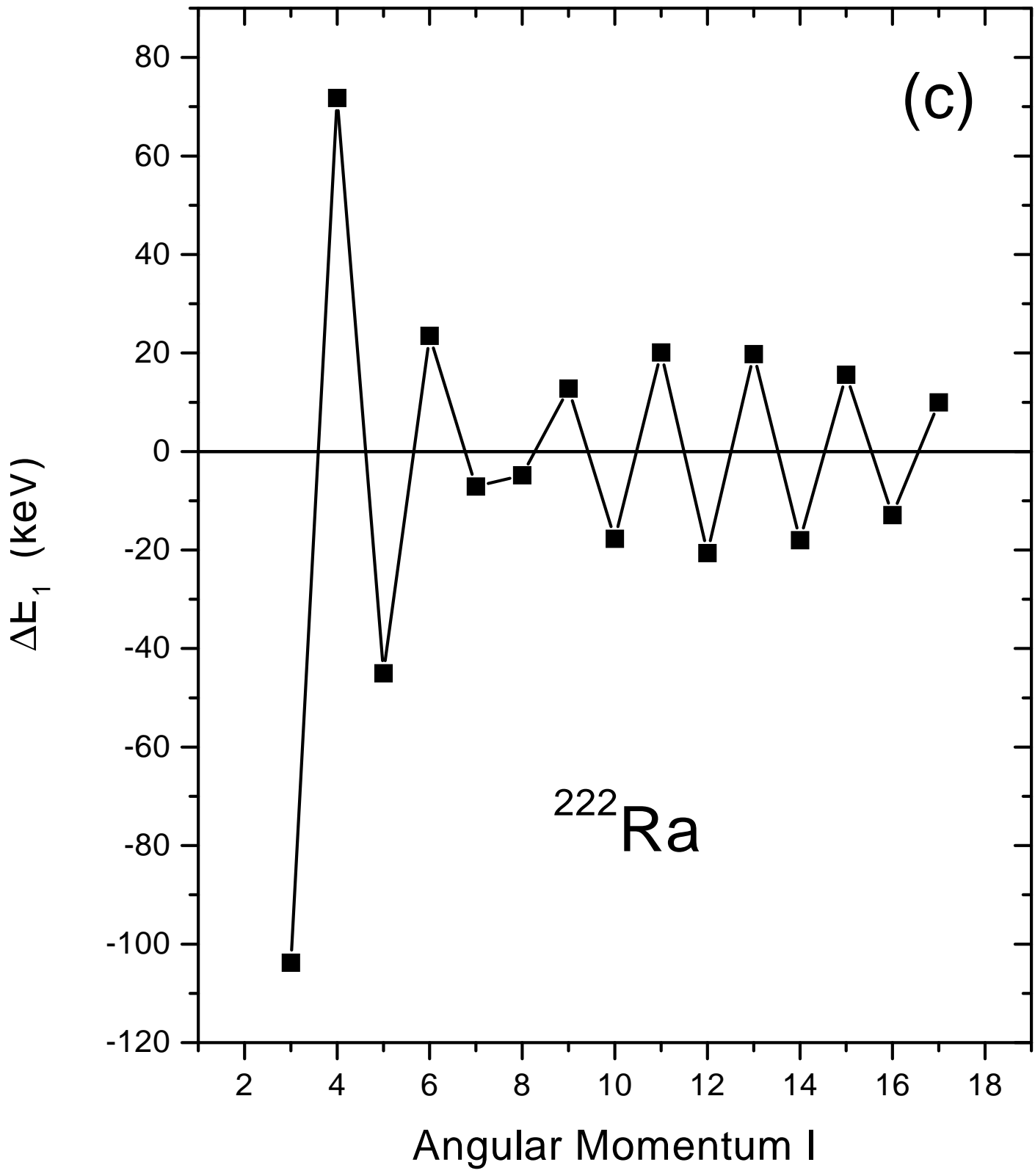
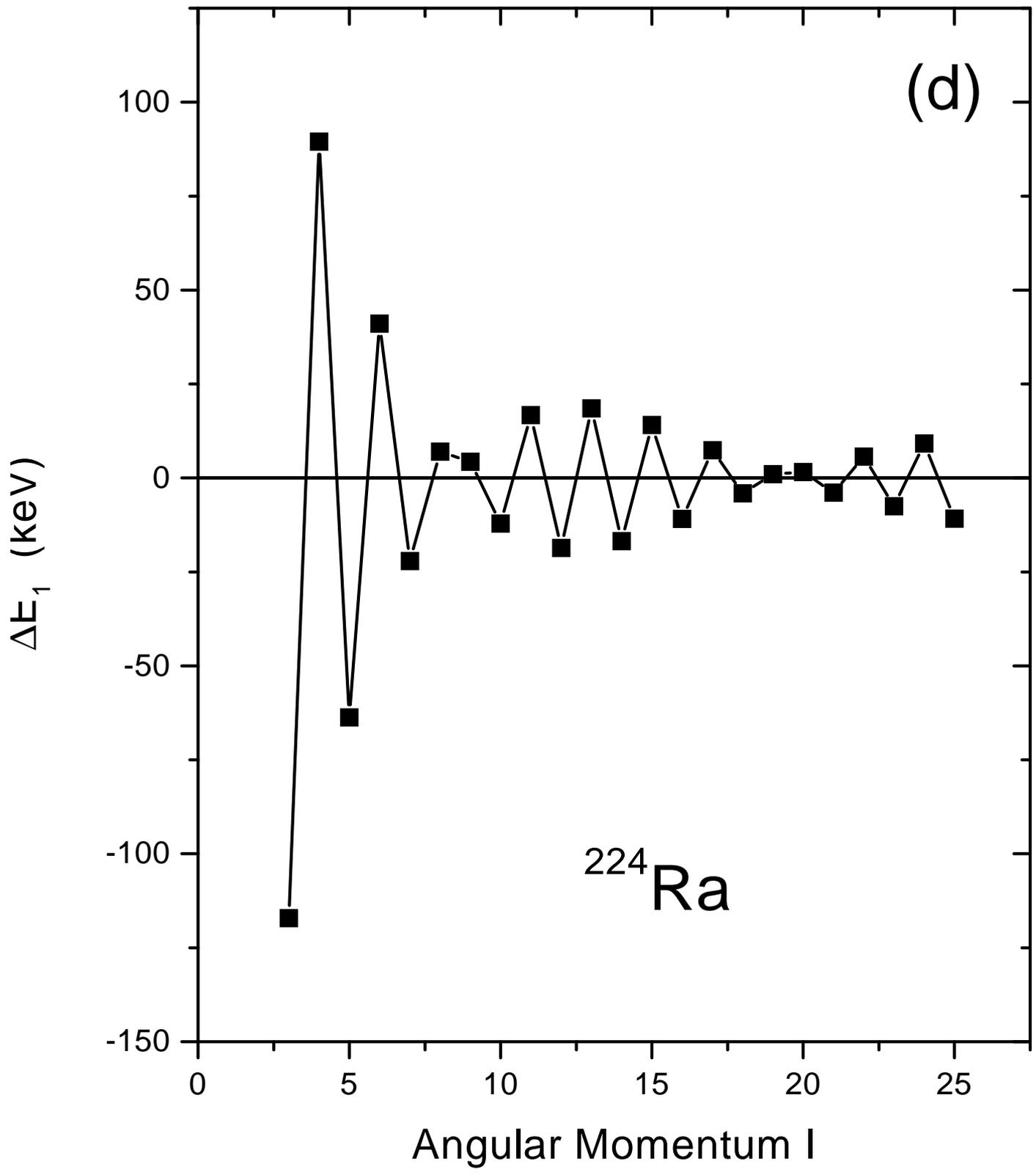
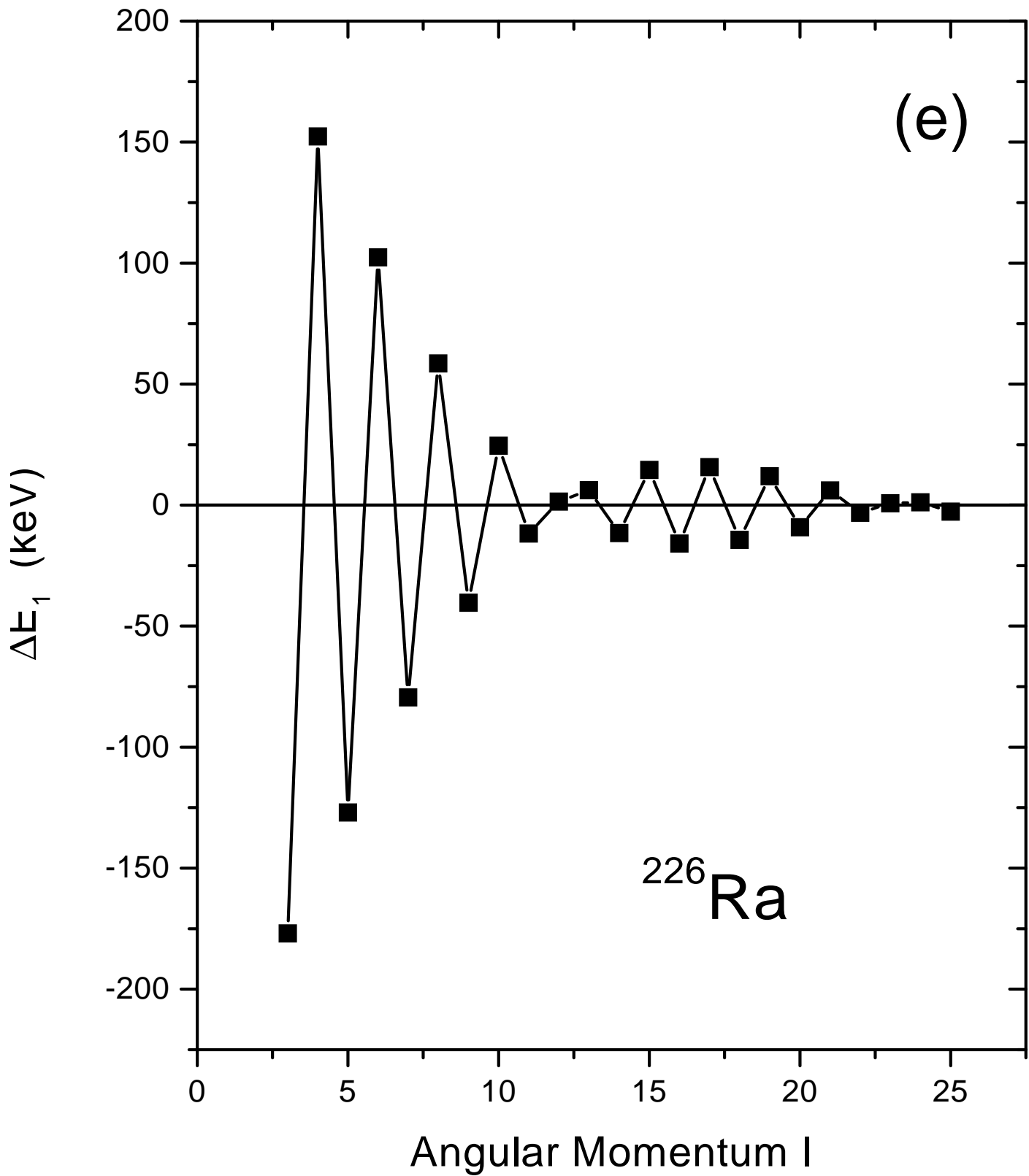
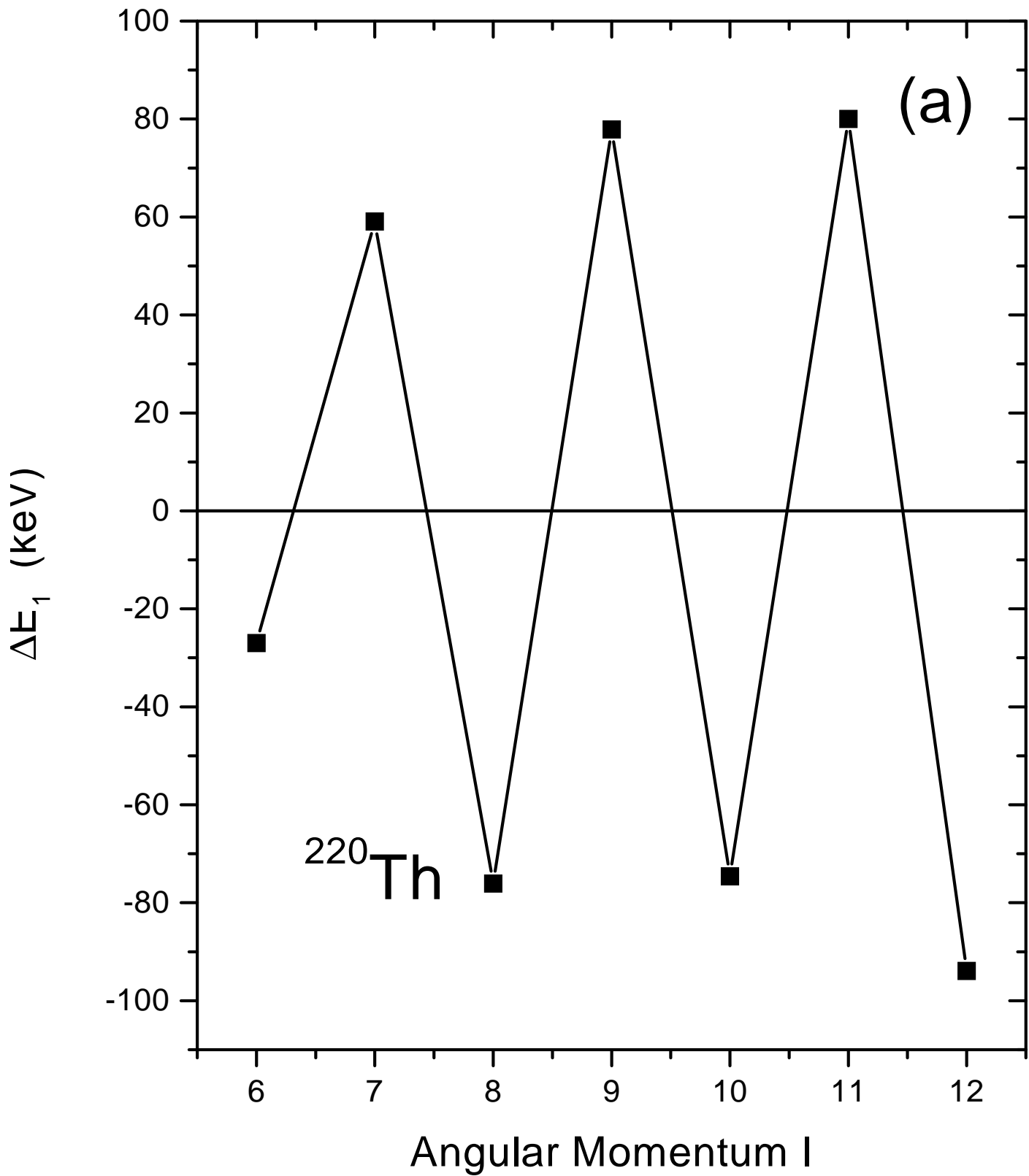
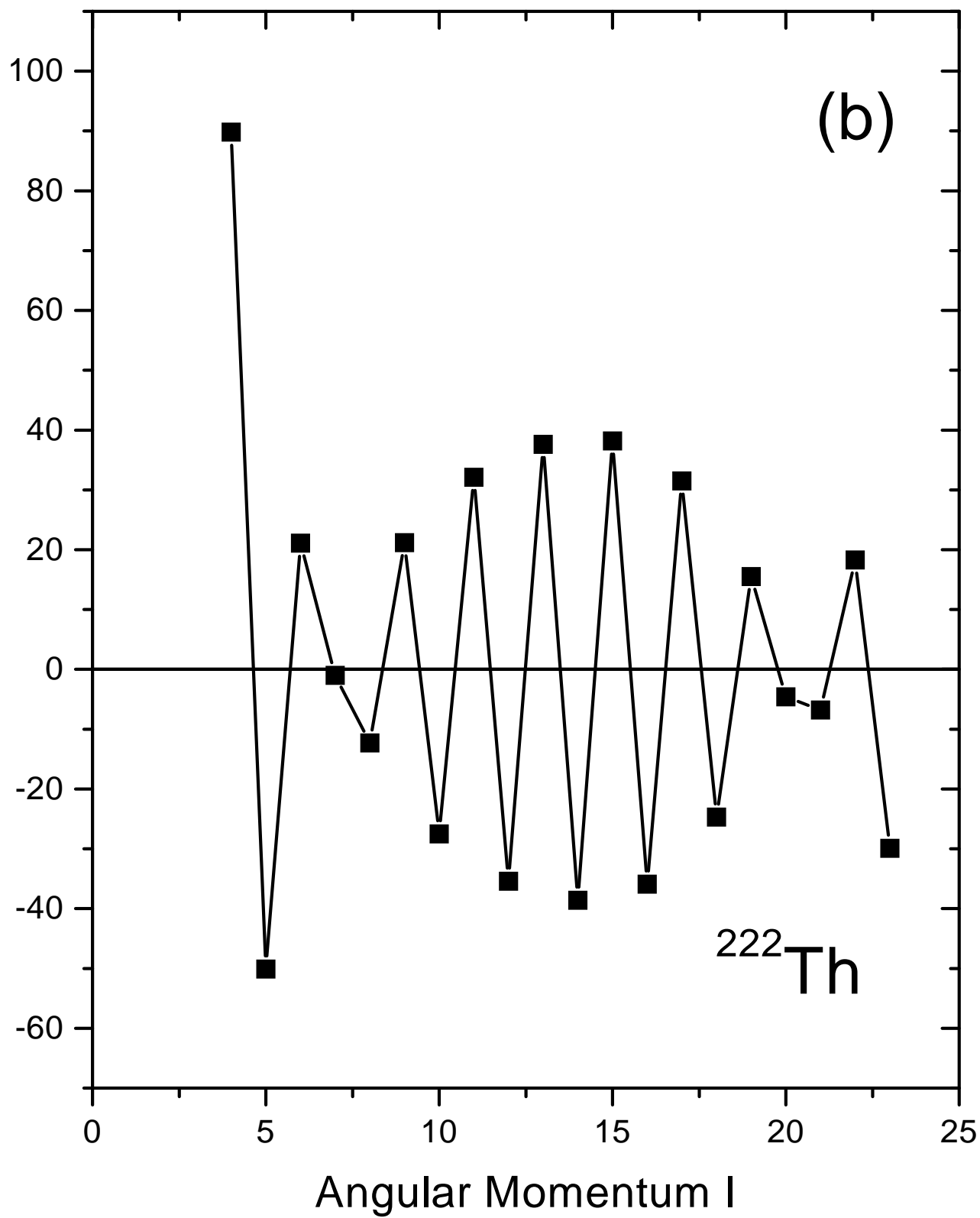
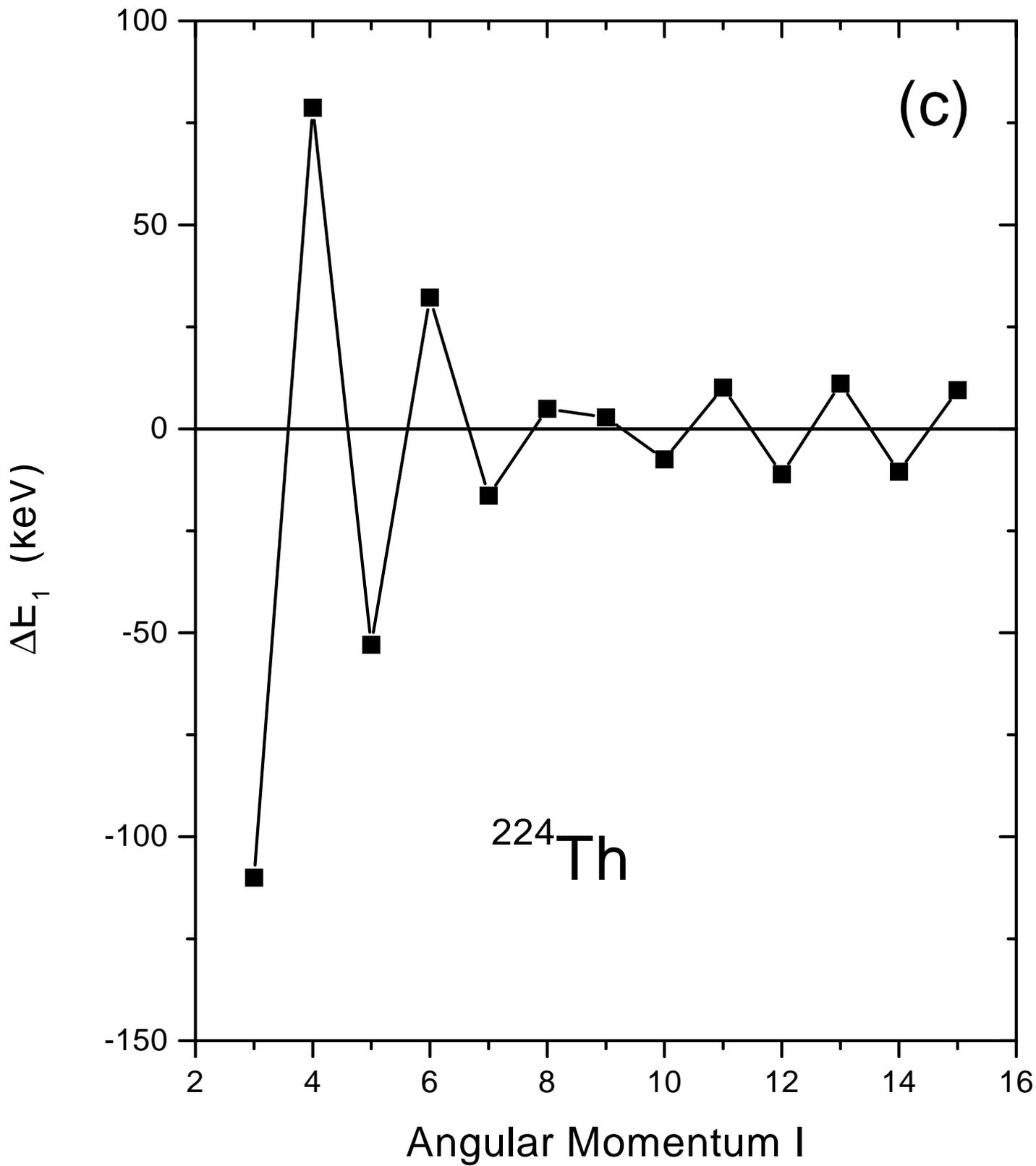
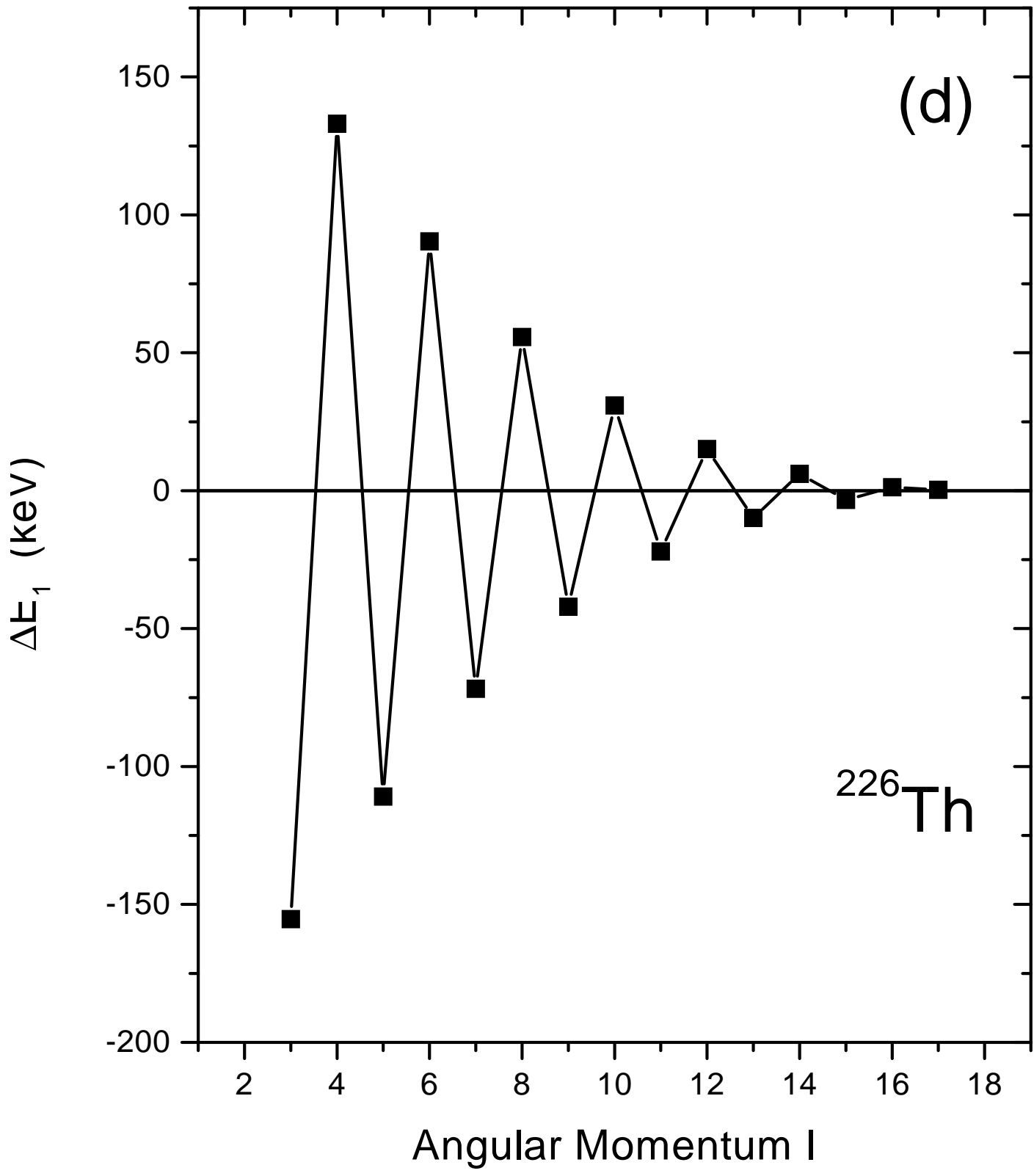
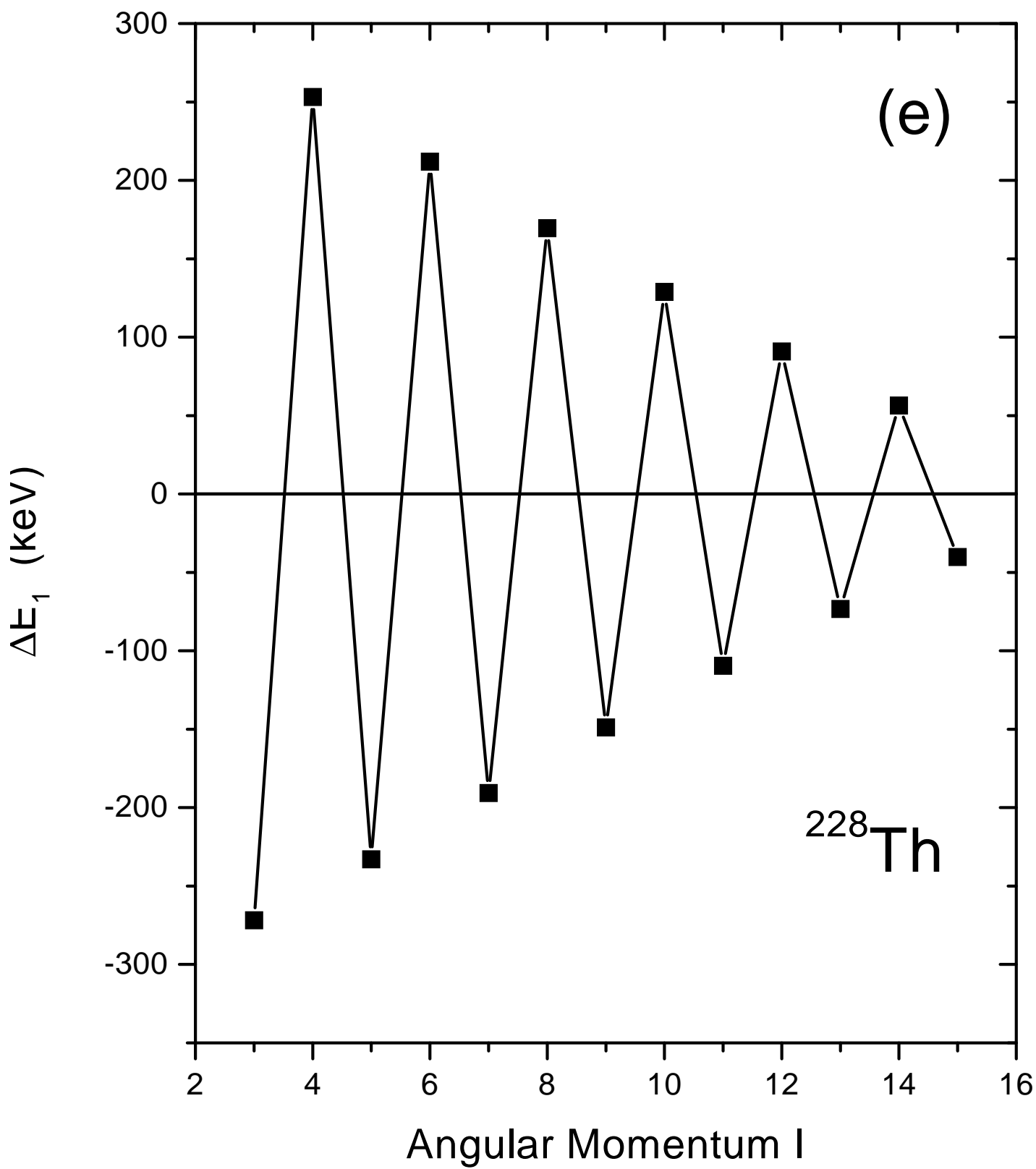
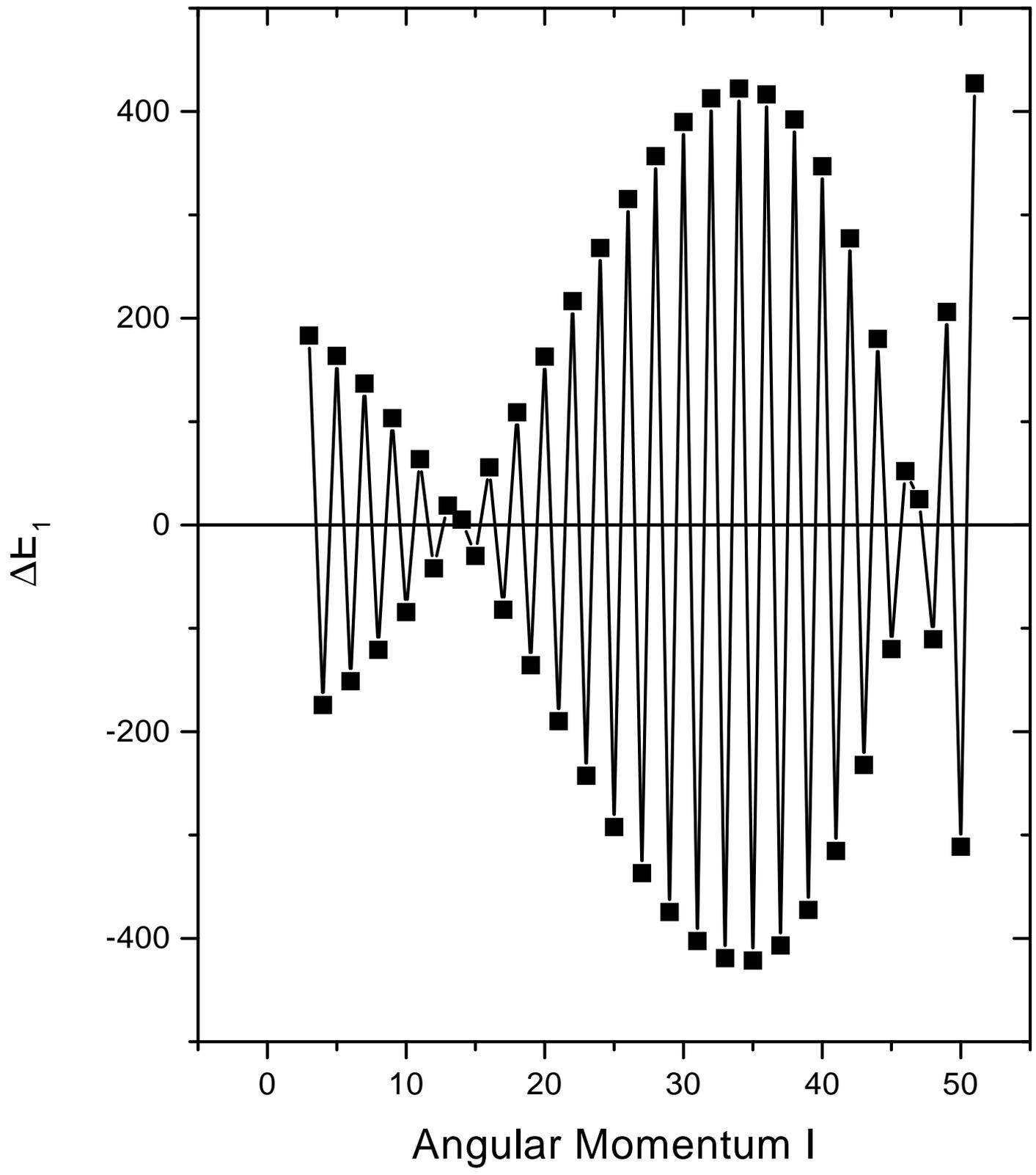
\end{document}